# Polarization-sensitive terahertz time-domain spectroscopy system without mechanical moving parts

*Mayuri Nakagawa[a]), Makoto Okano [a,b]), and Shinichi Watanabe [a,\*])*

*a) Department of Physics, Faculty of Science and Technology, Keio University, 3-14-1 Hiyoshi, Kohoku-ku, Yokohama, Kanagawa 223-8522, Japan*

*b) National Defense Academy, 1-10-20 Hashirimizu, Yokosuka-shi, Kanagawa 239-8686, Japan*

\*Corresponding author: watanabe@phys.keio.ac.jp

## Abstract

We report on the measurement of terahertz electric-field vector waveforms by using a system that contains no mechanical moving parts. It is known that two phase-locked femtosecond lasers with different repetition rates can be used to perform time-domain spectroscopy without using a mechanical delay stage. Furthermore, an electro-optic modulator can be used to perform polarization measurements without rotating any polarizers or waveplates. We experimentally demonstrate the combination of these two methods and explain the analysis of data obtained by such a system. Such a system provides a robust platform that can promote the usage of polarization-sensitive terahertz time-domain spectroscopy in basic science and practical applications. For the experimental demonstration, we alter the polarization of a terahertz wave by a polarizer.

## I. Introduction

Since the invention of terahertz time-domain spectroscopy (THz-TDS), terahertz spectroscopy has become an important tool for investigations of low-frequency optical responses in condensed matter physics [1]. In particular, polarization-sensitive (PS) THz-TDS has been widely applied in the field of material science; this technique can be used to probe a vast range of low-frequency anisotropic optical responses such as the anisotropic conductivity of carbon nanotubes [2], birefringence of polymers [3], anisotropic phonon absorption [4,5], cyclotron resonances [6-8], and chirality of metamaterials and biomolecules [9]. In addition, PS THz-TDS can be used for various practical and industrial applications such as non-contact characterization of semiconductors and polymeric materials [10]. Considering the wide applicability, it is desirable to construct a PS THz-TDS system that is less affected by environmental conditions than conventional PS THz-TDS systems.

In general, two femtosecond laser pulses emitted from a single light source and divided by a beam splitter are used for the terahertz pulse generation and detection in THz-TDS. The first pulse excites a photoconductive antenna or an electro-optic (EO) crystal to generate a terahertz pulse, and the second pulse is used to probe the electric-field (E-field) of the terahertz pulse that has passed through the sample. The probe pulse is appropriately delayed to determine the instantaneous field amplitude of the terahertz pulse, for instance, by using the Pockels effect. In this method, the polarization of the probe pulse is rotated in another crystal (such as ZnTe) by the terahertz-E-field-induced Pockels effect according to the temporal overlap between the two pulses, and then the polarization rotation is measured by a photodetector. Usually, to record the whole temporal profile of the terahertz E-field, a mechanical delay stage is used because it allows us to directly control the relative timing between the terahertz and the probe pulses. To enable polarization-sensitive measurements, an additional system component is required. In general, PS THz-TDS is performed by placing a wire-grid polarizer (WGP) in the terahertz beam path and investigating the polarization of the terahertz pulse after the sample by manually rotating the polarizer [11]. Other polarization measurement methods, such as rotating the EO crystal, have also been reported [12]. It has been reported that polarization modulation techniques based on mechanically rotating a terahertz waveplate [13], a WGP [14], or an EO crystal [15,16] can be used to improve the measurement speed and accuracy of PS THz-TDS. Furthermore, there are even modulation techniques that do not require mechanical rotating components: techniques based on the polarization modulation of the probe pulse by a photoelastic modulator [17] or an electro-optic modulator (EOM) [18] have also been reported. These modulation techniques aim at polarization measurements whose precision is not restricted by the stability of mechanical apparatuses. Although such sophisticated polarization measurement systems have been realized, a mechanical delay stage is still required in the conventional PS THz-TDS implementation scheme, which limits the measurement speed and robustness.

In 2005, an attractive alternative to the conventional THz-TDS implementation scheme was proposed and realized [19,20]. The proposed alternative time-domain measurement method, the so-called asynchronous optical sampling (ASOPS) method, requires no mechanical delay stage because two femtosecond laser pulses emitted from two different laser light sources with slightly different repetition rates are used for the terahertz pulse generation and detection, and the terahertz E-field transient is directly obtained from the temporal electric response of the photodetector. The ASOPS method has several advantages compared to the conventional system. Firstly, the measurement speed is generally much faster than that of the conventional method [19]. Secondly, the frequency resolution is much higher [20]. Finally, the stable alignment of the terahertz and probe pulses in the ASOPS method improves the stability of the measurement [21]. Note that ultra-high resolution Doppler-limited THz-TDS spectroscopy has recently been demonstrated [22]. However, an ASOPS-based PS THz-TDS system has not yet been reported.

In principle, the combination of the ASOPS method and one of the above-mentioned polarization modulation methods can be used to realize a PS THz-TDS system without any mechanical moving parts, which should result in a robust terahertz polarization spectroscopy system with significant improvements in the measurement time. In this article, we report on such an ASOPS-based PS THz-TDS system. By using two frequency-comb lasers with frequency stabilization and an EOM, we were able to perform PS THz-TDS measurements without using a mechanical delay stage and without using a mechanical rotation of optical elements such as polarizers. We first describe the analytical procedure to interpret the measured data, and then we demonstrate measurements of the polarization state for a terahertz wave that passed through a WGP as an example, which proves the feasibility of our method. A statistical analysis of the polarization spectrum is also provided.

**II. Experimental setup and measurement principle**

Figure 1 shows the experimental setup of the ASOPS-based PS THz-TDS system. The green shaded area visualizes the terahertz beam. The second WGP in the figure, WGP2, is mounted on a rotation mount to mimic different samples. First, we describe the details of the EOM-based terahertz polarization measurement system [18]. As shown in Fig. 1, we use the femtosecond fiber laser labelled "Laser 2" (wavelength $\lambda \approx 1550$ nm) for the terahertz pulse generation; the terahertz pulses are generated by exciting a photoconductive antenna (THz-P-Tx, Fraunhofer HHI) biased at a voltage of 59 V. The generated terahertz pulses are guided by two parabolic mirrors (PMs) and are focused on a <111>-oriented zinc-telluride (ZnTe) crystal with a thickness of 1 mm. The transmission axis of the first WGP in the terahertz beam path (WGP1) is fixed and parallel to the *Y*-axis, which is perpendicular to the optical table. The E-field vector transient of the terahertz pulse in the ZnTe crystal is measured by analyzing the polarization rotation of the probe light due to the terahertz-field-induced Pockels effect in the ZnTe crystal. The probe light is generated from the 1550-nm light of the femtosecond fiber laser labelled "Laser 1" by the second-harmonic generation (SHG) inside a periodically poled lithium niobate (PPLN) crystal. The polarization rotation of the probe pulse in the ZnTe crystal is analyzed by a combination of a quarter-wave plate (Fig. 1; "λ/4" after the ZnTe crystal), an EOM, a Wollaston prism (WP), and a balanced photodetector (BPD). To analyze the polarization of the terahertz E-field, we perform the rapid polarization analysis described in [18]: by using an EOM, it is possible to determine both the amplitude and direction of the terahertz E-field without any mechanical movements of optical components, resulting in a rapid and robust determination of the polarization of the terahertz pulse. The expression derived for the measured electric signal $\Delta I$ in this modulation method is provided in Eq. (1) (see Appendix for details), where the terahertz E-field vector probed at the delay time $t$ is defined as $\vec{E}_{\text{THz}}(t)$ [18]:

$$\Delta I(\tau) \approx C' \cdot |\vec{E}_{\text{THz}}(t)| J_0(m_f) \sin(\theta(t) - 3\phi)$$

$$+ 2|\vec{E}_{\text{THz}}(t)| \sum_{k=1}^{\infty} [J_{2k}(m_f) \sin(\theta(t) - 3\phi) \cos 2k\omega_M \tau$$

$$+ J_{2k-1}(m_f) \cos(\theta(t) - 3\phi) \sin(2k-1)\omega_M \tau]. \quad (1)$$

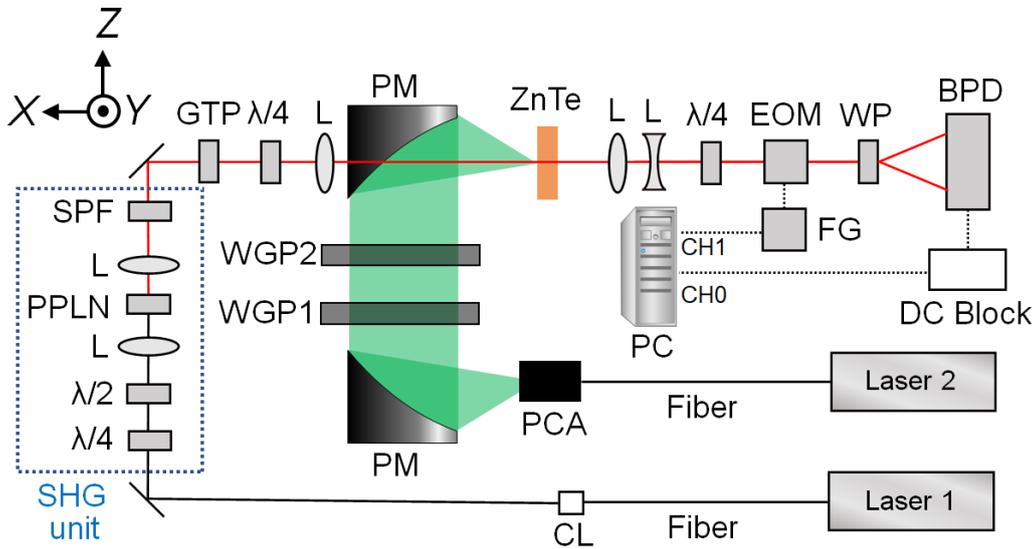

Fig. 1. Schematic of the experimental setup. The two femtosecond fiber lasers are indicated on the right lower side of the figure. The components used in the setup are as follows: CL: collimator lens, λ/4: quarter-wave plate, λ/2: half-wave plate, L: lens, PPLN: periodically poled lithium niobate crystal, SPF: short-pass filter, GTP: Glan–Thompson polarizer, EOM: electro-optic modulator, WP: Wollaston prism, BPD: balanced photodetector, DC Block: electronic DC block filter, FG: function generator, PC: personal computer, PCA: photoconductive antenna, PM: parabolic mirror, WGP1 and WGP2: 1st and 2nd wire-grid polarizer, ZnTe: zinc-telluride crystal. A digitizer in a PC records both the signal from the BPD (CH0: channel 0) and the voltage applied to the EOM (CH1: channel 1).

The definitions of the variables in Eq. (1) are as follows: $C'$ is a constant. $\tau$ is the laboratory time (for example, the actual temporal variation of the voltage applied to the EOM and the temporal overlap between the terahertz pulse and the probe pulse are described in terms of the laboratory time). $\theta(t)$ is the angle between the direction of $\vec{E}_{\text{THz}}(t)$ and the X-axis, and $\phi$ is the angle between the $[2\bar{1}\bar{1}]$ direction of the ZnTe crystal and the X-axis. The angular frequency $\omega_M$ describes the modulation frequency of the voltage applied to the EOM, which is synchronized with the difference between the repetition rates of Laser 1 and Laser 2 as described later. $m_f = \pi (V_0/V_\pi)$ where $V_0$ is the amplitude of the applied voltage and $V_\pi$ is the half-wave voltage of the EOM. $J_k$ ($k = 0,1,2, …$) are the Bessel functions of the first kind of order $k$. In our experiment, $V_\pi = 265$ V and $V_0 = 173$ V. For the analysis of the E-field transient, we simultaneously recorded the electric signal $\Delta I(\tau)$ (on channel 0) and the voltage applied to the EOM, $V_{\text{EOM}}(\tau)$, (on channel 1) with a sampling frequency of $f_1$ using a digitizer (M2p.5962-x4, Spectrum) in a personal computer. The time interval between the sampling data points is thus $\Delta\tau = 1/f_1$. The voltage signal $V_{EOM}(\tau)$ is required to determine the amplitude and phase of the signal modulation by the EOM with respect to the terahertz signal for the polarization analysis.

Next, we describe the details of the ASOPS system used in this study, which is based on a dual-comb spectroscopy system [23]. The two femtosecond fiber lasers have slightly different repetition rates ($f_1$ and $f_2$) with a frequency difference $\Delta f = f_2 - f_1$. The frequency stabilization of $f_1$ and $f_2$ is achieved by the frequency-locking scheme of dual-comb spectroscopy [24]: $f_1$ is phase-locked to a radio-frequency (RF) reference signal generated by a function generator referenced to a global-positioning-system-controlled rubidium (Rb) clock. The carrier envelope offset frequency of Laser 1 ($f_{ceo1}$) is phase-locked to another RF reference signal, thus Laser 1 works as a frequency comb [25] because the optical spectrum consists of equidistant frequency components with definite absolute frequency referenced to the Rb clock. Our system also contains a reference continuous-wave (CW) laser with frequency $f_{CW}$. It is phase-locked to one of the comb lines (with mode number $n$) of Laser 1 with the optical beat frequency $f_{beat1}$ to fulfill the condition $f_{CW} = f_{ceo1} + n \times f_1 + f_{beat1}$. The carrier envelope offset frequency of Laser 2 ($f_{ceo2}$) and the optical beat frequency ($f_{beat2}$) between $f_{CW}$ and one of the comb lines of Laser 2 (with mode number $n'$) are also phase-locked to RF reference signals to achieve $f_{CW} = f_{ceo2} + n' \times f_2 + f_{beat2}$. $f_2$ is stabilized through the above phase-locking scheme; $f_2 =$

( $f_{ceo1}$ - $f_{ceo2}$ + $n \times f_1$ + $f_{beat1}$ - $f_{beat2}$ ) / $n'$. The method to determine the two integers $n$ and $n'$ is described in [26]. To realize the coherent averaging condition [28], it is necessary to choose the frequencies of the RF reference signals, $f_{ceo1}$ and $f_{ceo2}$, in such a way that the ratio between $f_1$ and $\Delta f$, defined as $M \equiv f_1/\Delta f$, becomes an integer value. Here, we set $f_1$ = 61.530691 MHz and $M$ = 431,455, resulting in $f_2 \approx$ 61.530833612 MHz, and $\Delta f \approx$ 142.612 Hz.

The essential point of the present study is the combination of the ASOPS and EOM methods to realize a PS THz-TDS system with no mechanical moving parts. Below, we show a mathematical representation of the measured signal obtained by the combination of the ASOPS and EOM methods. First, we consider the situation without polarization modulation for simplicity. In the conventional THz-TDS implementation scheme, a mechanical delay stage is used to change the probe-pulse delay time $t$. On the other hand, in the ASOPS method, the delay time is automatically changed from pulse to pulse owing to the difference between the repetition rates of the terahertz pulse ($f_2$) and that of the probe pulse ($f_1$). Due to the different repetition rates, the temporal overlap between the terahertz pulse and the probe pulse at the ZnTe crystal is different for consecutive probe pulses, but after a temporal interval of $1/\Delta f$ in the laboratory time $\tau$, the same delay-time condition for the measurement of $E_{THz}(t)$ is realized again. Here, we measured the signal $\Delta I(\tau)$ with a sampling frequency $f_1$ by a digitizer. Because we set the ratio of $f_1$ to $\Delta f$ to the integer value $M$, the terahertz E-field transient are repeatedly recorded with a period of $M$ sampling points. The total temporal window of the terahertz transient that can be scanned by the probe pulses in our system is $1/f_2$. Therefore, the incremental change in the delay time $t$ for consecutive probe pulses is $\Delta t = 1/(f_2 \times M) = \Delta f/(f_2 \times f_1)$. Since $\Delta\tau = 1/f_1$, the relation $\Delta\tau = (M+1) \times \Delta t$ holds. From this relation, the relation between the laboratory time and delay time can be written as

$$\tau = (M + 1) \times t. \quad (2)$$

Equation (2) shows that the measured time-domain signal $\Delta I(\tau)$ automatically reflects the actual time-domain waveform of $E_{\text{THz}}$ with a magnified time scale. Equation (2) can be rewritten in terms of $\Delta f$ and $f_2$:

$$f_2 t = \Delta f \tau. \quad (3)$$

Next, we consider the situation when polarization modulation by the EOM is additionally performed to determine the terahertz E-field vector transient. Note that the measured terahertz E-field transient is a sum of Fourier components with equidistant frequency components at integer multiples of $f_2$. Therefore, we can define two orthogonal components of the E-field vector time-domain waveform as follows:

$$E_{\cos}(t) \equiv |\vec{E}_{\text{THz}}(t)| \cdot \cos(\theta(t) - 3\phi) = \sum_{m=-(M-1)/2}^{(M-1)/2} E_{m,\cos} \cdot e^{2\pi i m f_2 t}, \quad (4)$$

$$E_{\sin}(t) \equiv |\vec{E}_{\text{THz}}(t)| \cdot \sin(\theta(t) - 3\phi) = \sum_{m=-(M-1)/2}^{(M-1)/2} E_{m,\sin} \cdot e^{2\pi i m f_2 t}, \quad (5)$$

where $E_{m,\cos}$ and $E_{m,\sin}$ are the Fourier components at frequency $mf_2$. In this study, the modulation frequency of the EOM is set to $\omega_M = 2\pi(\Delta f/8)$, which is synchronized with $\Delta f$. By using Eqs. (3)–(5) and $\omega_M = 2\pi(\Delta f/8)$, Eq. (1) can be rewritten in terms of $E_{m,\cos}$, $E_{m,\sin}$, $J_k$, $M$, $\Delta f$, and $\tau$,

$$\Delta I(\tau) \propto J_0(m_f) \sum_{m=-(M-1)/2}^{(M-1)/2} E_{m,\sin} \cdot e^{2\pi i m \Delta f \tau}$$

$$+ \sum_{k=1}^{\infty} \left[ \sum_{m=-(M-1)/2}^{(M-1)/2} J_{2k}(m_f) E_{m,\sin} \cdot \left( e^{2\pi i \left(m+\frac{k}{4}\right)\Delta f \tau} + e^{2\pi i \left(m-\frac{k}{4}\right)\Delta f \tau} \right) - i \cdot J_{2k-1}(m_f) E_{m,\cos} \right.$$

$$\left. \cdot \left( e^{2\pi i \left(m+\frac{(2k-1)}{8}\right)\Delta f \tau} - e^{2\pi i \left(m-\frac{(2k-1)}{8}\right)\Delta f \tau} \right) \right]. \quad (6)$$

In contrast to the ASOPS-based THz-TDS method, where the same measurement condition is reached at laboratory-time intervals of $1/\Delta f$, our ASOPS-based PS THz-TDS system reaches the same delay-time condition for the measurement of $\vec{E}_{\text{THz}}(t)$ after a time interval of $8/\Delta f$. Equation (6) shows that the frequency sideband component at $\left(m \pm \frac{1}{4}\right)\Delta f$ $\left[\left(m \pm \frac{1}{8}\right)\Delta f\right]$ of the measured signal corresponds to the amplitude of the sine [cosine] component at frequency $mf_2$ of the terahertz E-field transient, $E_{m,\sin}$ $[E_{m,\cos}]$, multiplied by $J_2(m_f)$ $[J_1(m_f)]$. Thus, we need to perform a Fourier analysis of the measured $\Delta I(\tau)$ signal to derive $E_{m,\sin}$ and $E_{m,\cos}$. In this analysis, we only use terms proportional to $J_2(m_f)$ and $J_1(m_f)$ (corresponding to $k=1$ in the sum). Note that other sidebands of different terahertz frequency components appear at the same frequency [27]. For instance, the sideband corresponding to $k = $ 3 of the $(m+1)\Delta f$-frequency component appears at the frequency of $\left(m+1-\frac{3}{4}\right)\Delta f = \left(m+\frac{1}{4}\right)\Delta f$ with the amplitude $J_6(m_f)$, which spectrally overlaps with the sideband corresponding to $k=1$ of the $m\Delta f$-frequency component with the amplitude $J_2(m_f)$. However, we neglect the former term because $J_6(m_f) \ll J_2(m_f)$.

Another analytical representation that is useful for time-domain data analysis is obtained when we consider $\Delta I(\tau)$ at different laboratory times $\tau = \tau_0 + \frac{l}{\Delta f}$, where $\tau_0$ is a specific laboratory time and $l$ is an integer value ranging from 0 to 7. The corresponding measurement conditions provide results for $\vec{E}_{\text{THz}}(t)$ at $t = t_0 + \frac{l}{f_2}$ where $t_0 = (M+1)^{-1} \cdot \tau_0$. Because the repetition frequency of the terahertz pulse train is $f_2$, the relation $\vec{E}_{\text{THz}}\left(t_0 + \frac{l}{f_2}\right) = \vec{E}_{\text{THz}}(t_0)$ holds. Therefore, we can derive the following relation from Eq. (1):

$$\Delta I\left(\tau_0 + \frac{l}{\Delta f}\right) \propto |\vec{E}_{\text{THz}}(t_0)| J_0(m_f) \sin(\theta(t_0) - 3\phi)$$

$$+ 2|\vec{E}_{\text{THz}}(t_0)| \sum_{k=1}^{\infty} \left[ J_{2k}(m_f) \sin(\theta(t_0) - 3\phi) \cos 2k\omega_M \left(\tau_0 + \frac{l}{\Delta f}\right) \right.$$

$$\left. + J_{2k-1}(m_f) \cos(\theta(t_0) - 3\phi) \sin(2k-1)\omega_M \left(\tau_0 + \frac{l}{\Delta f}\right) \right]. \quad (7)$$

Equation (7) can be used instead of Eq. (6) to retrieve the terahertz E-field vector at a specific time $t_0$ [$\vec{E}_{\text{THz}}(t_0)$] by analyzing the eight values of $\Delta I(\tau)$ at $\tau = \tau_0 + \frac{l}{\Delta f}$ with $l = 0, 1, 2, \cdots, 7$. These values of $\Delta I(\tau)$ reflect different modulation amplitudes induced by the EOM.

## III. Experimental setup and measurement principle

In this section, we show several experimental results obtained by combining the ASOPS and EOM methods. First, we show measured raw data and explain how the polarization state of the terahertz waves can be determined. Then, we show results of polarization measurements for a terahertz wave transmitted through a polarizer using different polarization angles to verify our method.

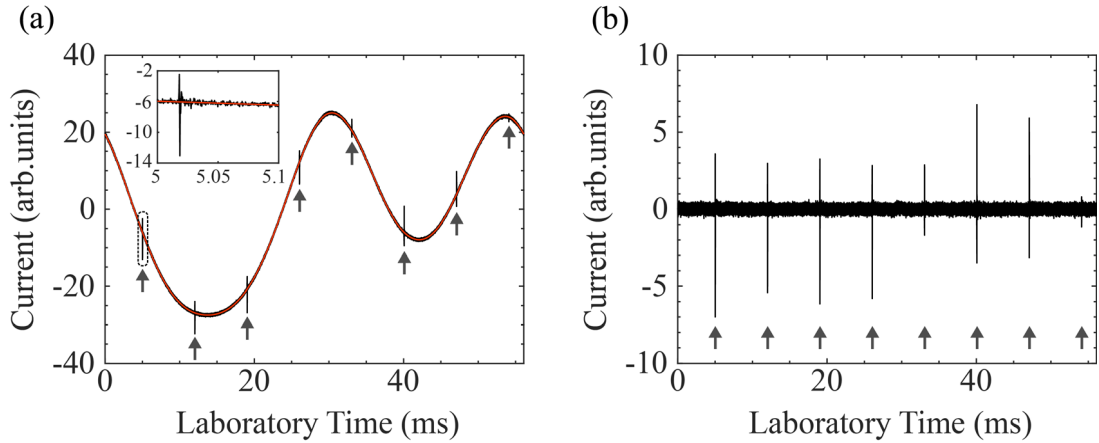

Fig. 2. (a) Current signal recorded on channel 0 after averaging (black line). The integration time for averaging was about 5 minutes. The red curve is the result of the fit for the baseline subtraction. The arrows indicate the times when sharp spikes appear in the signal. (Inset: Magnification of the data around the first spike at τ ≈ 5 ms). (b) The "flattened" signal (after the baseline subtraction).

*3.1. Analysis of raw data*

Figure 2(a) shows the average of the current signal recorded on channel 0 when the angles of the transmission axes of WGP1 and WGP2 were 90° and 130° relative to the *X*-axis, respectively. This current signal is proportional to $\Delta I$. As we set $\Delta f \approx 142.612$ Hz and $\omega_M = 2\pi(\Delta f/8)$, the period of the time-domain current signal for the polarization measurement is $8/\Delta f \approx 56$ ms in the laboratory time. The measured transient in the range $8/\Delta f$ contains $8 \times M = 3{,}451{,}640$ data points, and we integrated the data for about 5 minutes; we averaged ~5,300 individual transients to improve the signal to noise ratio. There are eight sharp spikes in the averaged transient in Fig. 2(a), which are indicated by the arrows. The inset of Fig. 2(a) is a magnification of the data around the first spike at $\tau \approx 5$ ms and clarifies that these spikes are the peak structures of the terahertz time-domain waveform, which are repeated with a period of $1/\Delta f$. In addition, we find that the recorded signal also contains a slowly oscillating component. This slow oscillation is due to a slight misalignment of the polarization components in the setup, which is difficult to remove by usual alignment procedures. However, because the frequency of this oscillation is much smaller than the frequency of the terahertz time-domain signal, we can easily subtract this component from the measured data. Here, we fitted the original data to a sum of ten sine functions with different frequencies and phases. The fitting result is indicated by the red curve in Fig. 2(a). All ten fitting frequencies are smaller than 23 kHz [corresponding to less than 10 GHz in the optical frequency range according to Eq. (2)], and thus are much smaller than the frequencies related to the terahertz signal. Figure 2(b) shows the "flattened" signal obtained by subtracting the slowly oscillating sine components from the original data in Fig. 2(a). The magnitudes of the eight spike structures in the flattened signal in Fig. 2(b) are not equal because of the polarization modulation of the probe beam by the EOM. We can retrieve the polarization information from the amplitude variation of these spikes. Thus, the polarization state of the terahertz wave is derived by substituting the signal in Fig. 2(b) for $\Delta I(\tau)$ in Eq. (6).

The numerical procedure to retrieve the terahertz time-domain E-field vector waveform is as follows. First, we calculate the Fourier transform of the signal in Fig. 2(b) and calculate $E_{m,\cos}$ and $E_{m,\sin}$ as explained in Section 2. Then, the two orthogonal components of the terahertz E-field vector, $E_{\cos}(t)$ and $E_{\sin}(t)$ are calculated using Eqs. (4) and (5). Figure 3(a) shows the three-dimensional plot of vector waveform composed of the two orthogonal components. However, as shown in Eqs. (4) and (5), the angle of the E-field vector retrieved from $E_{\cos}(t)$ and $E_{\sin}(t)$ is not $\theta(t)$ but $\alpha(t) = \theta(t) - 3\phi$, which is the angle between $\vec{E}_{\mathrm{THz}}(t)$ and a certain crystal direction of the ZnTe crystal [18]. In order to retrieve the actual angle with respect to the X-axis, $\theta(t)$, we performed an additional experiment to evaluate $\phi$; We changed the angle of the transmission axis of WGP2 to 90° relative to the X-axis (WGP1 was already parallel to the Y-axis), measured the data to obtain $E_{\cos}(t)$ and $E_{\sin}(t)$, and calculated the time-dependent angle of the E-field vector, $\alpha(t) = \tan^{-1}(E_{\sin}(t)/E_{\cos}(t))$. In this experimental configuration, $\theta(t=0)$ is either $90°$ or $-90°$, and the angle $\phi = (\theta(t=0) - \alpha(t=0))/3$. While there is an ambiguity in the sign of $\theta(t=0)$ in this calibration procedure, this ambiguity does not affect the determination of the azimuth and ellipticity angles of each frequency component of the E-field vector waveform. In Fig. 3(b), we plot the same data as in Fig. 3(a) around $t=0$ but added $3\phi$ to the angle $\alpha(t)$ to obtain $\theta(t)$. The terahertz E-field vector waveform is linearly polarized with an angle $\theta(t)$ ~ $130°$, consistent with the direction of the transmission axis of WGP2.

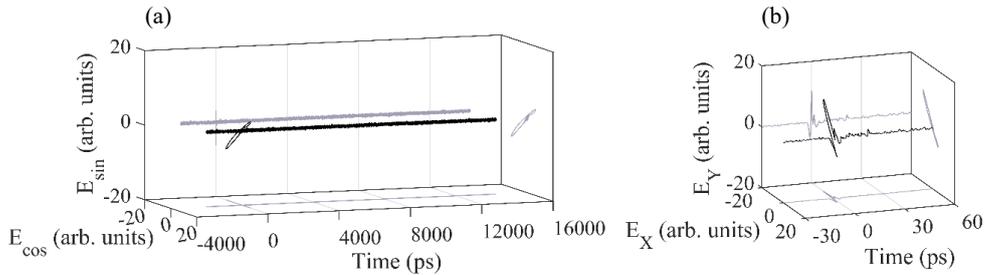

Fig. 3. Three-dimensional plot of the retrieved terahertz E-field vector waveform (a) before and (b) after the evaluation of the angle $\phi$. In (b), only the data around the peak structure is plotted. The retrieved polarization is consistent with the angle of the transmission axis of WGP2 ($\beta = -50° \pm 180°$).

*3.2. Measurement results for different angles of the WGP transmission axis*

To check the system performance, we show the results of polarization measurements for the terahertz wave transmitted through WGP2 using various angles of the transmission axes. The angle of the transmission axis of WGP2 with respect to the *X*-axis is hereafter denoted as *β*. The transmission axis of WGP1 was kept at 90° relative to the *X*-axis. Figure 4(a) shows the time-domain waveform of the *Y*-component of the terahertz E-field, $E_Y(t) \equiv |\vec{E}_{\text{THz}}(t)| \cos\theta(t)$, around *t* = 0 for *β* = 90°. Figure 4(b) shows the distribution of the spectral power density (SPD) obtained by the Fourier transform of the time-domain waveform. When we calculated the SPD spectra, we deleted the data outside the temporal window shown in Fig. 4(a) to improve the signal to noise ratio. Figure 4(c) shows the frequency-averaged SPD as a function of *β* (we averaged the SPD data over the range 0.10–0.75 THz where the signal intensity is sufficiently large). The data is well reproduced by a fitting function proportional to $\cos^2(\beta - 92.2°)$ as shown by the dashed curve. The result indicates that the transmission axis of WGP1 is tilted by about 2.2° relative to the transmission axis of WGP2 for *β* = 90°.

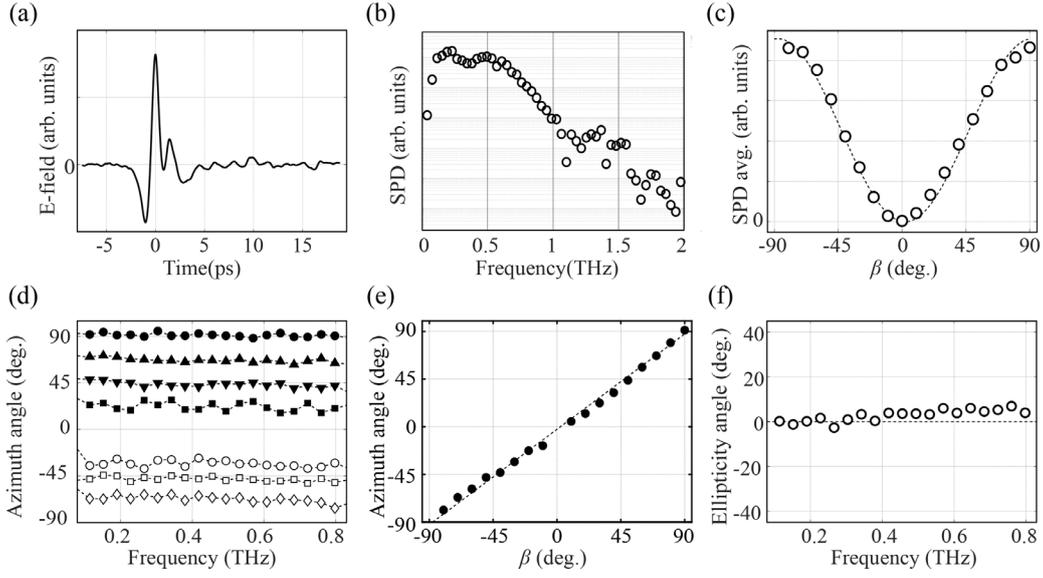

Fig. 4. (a) *Y*-component of the measured terahertz E-field vector waveform around the peak position obtained when the transmission axis of WGP2 was set to *β* = 90°. (b) Semilogarithmic plot of the distribution of the SPD calculated from (a). (c) Frequency-averaged SPD as a function of *β*. The dashed line is a fit to a squared cosine function. (d) Frequency dependence of the azimuth angle of the measured terahertz E-field vector waveform for seven different values of *β*: -70° (◊), -50° (□), -30° (○), 30° (■), 50° (▼), 70° (▲), and 90° (●). (e) Frequency-averaged azimuth angle as a function of *β*. The dashed line is a fit to a linear function. (f) Frequency dependence of the ellipticity angle obtained when *β* was set to 90°.

Figure 4(d) shows the spectra of the azimuth angle of the measured terahertz E-field vector waveform for seven different values of *β*. The spectra are almost independent of the frequency. In Fig. 4(e), we plot the frequency-averaged value of the azimuth angle as a function of *β* (averaging range: 0.10–0.75 THz; the data for *β* = 0 is not shown because the signal is too small to analyze the polarization state). The dashed curve is a fit of the data to a linear function, $\beta + c$, where the fitting parameter $c = -2.6°$. Although we performed the experiment to evaluate $\phi$ at $t = 0$ in advance to calibrate the angle of the transmission axis of WGP2 to be $\theta(t = 0) = 90°$ at $\beta = 90°$ in the calibration process, the frequency-averaged value of the azimuth angle at $\beta = 90°$ is 87.4° according to the fitting result. This discrepancy indicates that the azimuth angle derived from the frequency-domain analysis is slightly different from that evaluated at $t = 0$. This means that the azimuth angle of the terahertz wave varies temporally, i.e., the terahertz wave is elliptically polarized. Indeed, as shown in Fig. 4(f), the retrieved ellipticity angle deviates from 0° at higher frequencies. The observed ellipticity could be a consequence of using a parabolic mirror for focusing; we confirmed that a linearly polarized terahertz wave is accompanied by either a rotational or divergent electric field in the focal plane especially at higher frequencies when it is focused by a parabolic mirror [28,29], and the measured terahertz time-domain waveform is elliptically polarized when it is detected outside the exact focal point [30].

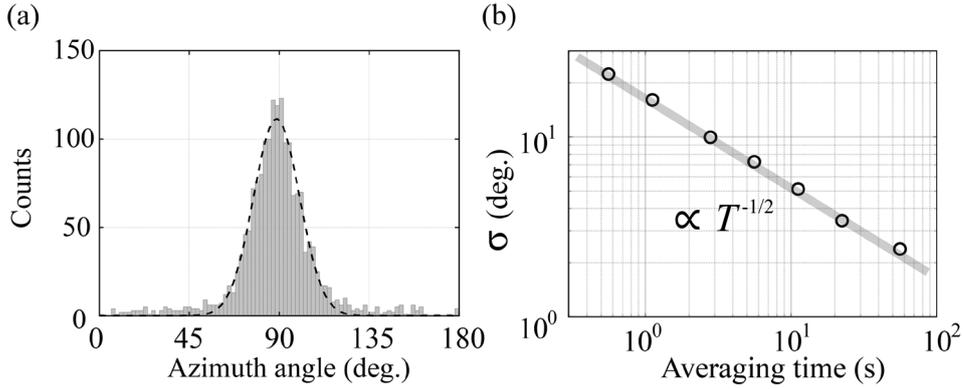

Fig. 5. (a) Statistical distribution of the frequency-averaged azimuth angles derived from the 1800 data segments of length $80/\Delta f$. (b) Standard deviation of σ as a function of the averaging time $T$. The gray line shows a $T^{-1/2}$ dependence.

Finally, we discuss statistical properties of the terahertz polarization spectrum measured by our system. For this discussion, the transmission axes of both WGP1 and WGP2 were set to 90° and we measured 18,000 EOM-modulated waveforms within 1,010 seconds. The data was divided into 1,800 data segments of equal length (segment length ≈ 560 ms, corresponding to 10 waveforms of length $8/\Delta f$), and then the average waveform of each data segment was calculated, resulting in a total of 1,800 averaged waveforms of length $8/\Delta f$. Additionally, to improve the signal-to-noise ratio for the frequency analysis, we deleted the time-domain data outside the temporal window of ±5 ps around the peak position. Then, the Fourier transform of each averaged waveform was performed, and we calculated the frequency dependence of the azimuth angle of each averaged waveform and averaged it over 0.10 to 0.75 THz. Figure 5(a) shows the statistical distribution of the frequency-averaged azimuth angle $\bar{\theta}$ for the 1800 average waveforms; $\bar{\theta}$ follows a normal distribution with a mean value of 89.0° and a standard deviation of $\sigma_0 = 11.5°$. As explained above, each data segment results in one value of $\bar{\theta}$. Each $\bar{\theta}$ value is thus result of a measurement with an averaging time of ≈ 560 ms. In Fig. 5(b), we plot the standard deviation of $\bar{\theta}$ as a function of the averaging time $T$. Here, the averaging time $T$ is an integral multiple of the length of one data segment, and as the number of considered data segments increases, the number of $\bar{\theta}$ values used to evaluate $\sigma$ increases. It is clearly found that $\sigma$ is inversely proportional to the square root of $T$. From this averaging-time dependence of $\sigma$, we determine that the azimuth angle is 89.0° ± 0.3° for $T$ = 1,010 s.

## IV. Conclusions

We have proposed and demonstrated an ASOPS-based PS THz-TDS system that contains no mechanical moving parts. The combination of the ASOPS and EOM methods enables us to measure terahertz E-field vector waveforms with a wide temporal range and short measurement time. A statistical analysis of the polarization spectrum has been presented to discuss the precision of the measured azimuth angle. In the field of materials science, the evaluation of the polarization state of a terahertz spectrum is important for the discussion of various anisotropic low-frequency responses. The proposed method enables us to quantitatively discuss the polarization change inside a material with a well-defined precision, which could promote such studies.

In this work, since we deleted the data points of the time-domain waveform that are not near the important signal features to improve the signal-to-noise ratio, only 0.06% of the total data were used for the spectrum analysis. This poor efficiency is due to the low repetition rates of the frequency-comb light sources, which results in long temporal windows of the ASOPS signal (here, $1/f_{rep}$ ~ 16 ns, while the important features of the terahertz time-domain signal are concentrated within only ~10 ps). To improve the measurement efficiency, it is important to use light sources with higher repetition rates. Recently, an ASOPS measurement at a repetition rate of 10 GHz has been reported [31]. In addition, high-speed fiber-coupled EOMs with several tens of GHz bandwidth are commercially available. A combination of high repetition rate lasers and high-speed EOMs may enable high-speed polarization-sensitive terahertz time-domain measurements with higher efficiencies in future.

**Appendix A: Jones matrix formulation of the polarization-sensitive terahertz E-field detection method using an EOM**

In this Appendix, we derive the expression for the signal obtained by the terahertz polarization measurement method based on EO crystals [32] using the Jones matrix formalism.

We consider the optical setup shown in Fig. 1: The light of Laser 1 is frequency doubled by a PPLN crystal and passes through a Glan–Thompson polarizer and a quarter-wave plate. After the quarter-wave plate, the probe pulse is circularly polarized, and thus its Jones vector is written as follows:

$$P_0 = \frac{1}{2}\begin{pmatrix} 1+i \\ 1-i \end{pmatrix}. \quad (A1)$$

When the terahertz pulse irradiates the ZnTe crystal, it causes birefringence in the crystal. The Jones matrix representation of the (111) ZnTe crystal including the effect of birefringence is

$$M_{\text{ZnTe}} = \begin{pmatrix} \cos\phi_2 & \sin\phi_2 \\ -\sin\phi_2 & \cos\phi_2 \end{pmatrix} \begin{pmatrix} e^{-\frac{iC}{2}} & 0 \\ 0 & e^{\frac{iC}{2}} \end{pmatrix} \begin{pmatrix} \cos\phi_2 & -\sin\phi_2 \\ \sin\phi_2 & \cos\phi_2 \end{pmatrix}, \quad (A2)$$

where [33]

$$C = -\frac{2\Omega L \chi^{(2)}}{c\sqrt{6\varepsilon_r}}|E_{\text{THz}}|, \quad (A3)$$

$$\phi_2 = -\frac{(\theta - 3\phi)}{2}. \quad (A4)$$

Here, $\Omega$ is the center wavelength of the probe pulse, $L$ is the thickness of the ZnTe crystal, $\chi^{(2)}$ is the second-order susceptibility, $\varepsilon_r$ is the relative permittivity of the ZnTe crystal, $\theta$ is the angle between the direction of $\vec{E}_{\text{THz}}(t)$ and the $X$-axis, and $\phi$ is the angle between the $[2\bar{1}\bar{1}]$ direction of the ZnTe crystal and the $X$-axis. To determine $\phi$, we performed a calibration procedure to determine the reference angle as described in Section 3.1. After the ZnTe crystal, the probe pulse passes through another quarter-wave plate and the EOM, whose Jones matrices are

$$M_{\lambda/4} = \frac{1}{\sqrt{2}}\begin{pmatrix} 1-i & 0 \\ 0 & 1+i \end{pmatrix}, \quad (A5)$$

and

$$M_{\text{EOM}} = \frac{1}{\sqrt{2}}\begin{pmatrix} 1 & 1 \\ -1 & 1 \end{pmatrix}\begin{pmatrix} e^{-\frac{i\varepsilon}{2}} & 0 \\ 0 & e^{\frac{i\varepsilon}{2}} \end{pmatrix}\frac{1}{\sqrt{2}}\begin{pmatrix} 1 & -1 \\ 1 & 1 \end{pmatrix} = \begin{pmatrix} \cos\frac{\varepsilon}{2} & i\sin\frac{\varepsilon}{2} \\ i\sin\frac{\varepsilon}{2} & \cos\frac{\varepsilon}{2} \end{pmatrix}, \quad (A6)$$

respectively, where $\varepsilon$ represents the voltage applied to the EOM with amplitude $m_f$, frequency $\omega_M$, and phase $\varphi_M$,

$$\varepsilon = m_f \sin(\omega_M \tau + \varphi_M). \quad (A7)$$

In the actual experiment, $m_f$, $\omega_M$, and $\varphi_M$ are monitored on channel 1 ($V_{\text{EOM}}$) as shown in Fig. 1(b). Therefore, we can shift the whole data to obtain the condition $\varphi_M = 0$. From Eqs. (A1), (A2), (A5), and (A6), we find that the polarization state of the probe pulse after the EOM can be written as

$$P_1 \equiv \begin{pmatrix} E_+ \\ E_- \end{pmatrix} = M_{\text{EOM}} \cdot M_{\lambda/4} \cdot M_{\text{ZnTe}} \cdot P_0$$

$$= \frac{1}{\sqrt{2}} \begin{pmatrix} \cos\frac{C}{2}\left(\cos\frac{\varepsilon}{2} + i\sin\frac{\varepsilon}{2}\right) + \sin\frac{C}{2}\left(\sin\left(2\phi_2 - \frac{\varepsilon}{2}\right) - i\cos\left(2\phi_2 - \frac{\varepsilon}{2}\right)\right) \\ \cos\frac{C}{2}\left(\cos\frac{\varepsilon}{2} + i\sin\frac{\varepsilon}{2}\right) + \sin\frac{C}{2}\left(-\sin\left(2\phi_2 - \frac{\varepsilon}{2}\right) + i\cos\left(2\phi_2 - \frac{\varepsilon}{2}\right)\right) \end{pmatrix}. \quad (A8)$$

Finally, the Wollaston prism splits the beam into the $E_+$ and $E_-$ components described in Eq. (A8), and the difference between the intensities of the two components, $\Delta I$, is measured by the balanced photodetector:

$$\Delta I = |E_+|^2 - |E_-|^2 = 2\left(\cos\frac{C}{2}\sin\frac{C}{2}\cos\frac{\varepsilon}{2}\sin\left(2\phi_2 - \frac{\varepsilon}{2}\right) - \cos\frac{C}{2}\sin\frac{C}{2}\sin\frac{\varepsilon}{2}\cos\left(2\phi_2 - \frac{\varepsilon}{2}\right)\right)$$

$$= \sin C \sin(2\phi_2 - \varepsilon). \quad (A9)$$

By substituting Eqs. (A3), (A4), and (A7) into Eq. (A9), and assuming $\sin C \approx C$, we obtain the following approximate expression for $\Delta I$:

$$\Delta I \approx -\frac{2\Omega L \chi^{(2)}}{c\sqrt{6\varepsilon_r}} |E_{\text{THz}}| \sin(-\theta + 3\phi - m_f \sin\omega_M \tau)$$

$$\propto |E_{\text{THz}}| \sin(\theta - 3\phi + m_f \sin\omega_M \tau). \quad (A10)$$

Equation (1) is derived from Eq. (A10) by a Fourier expansion.

**Funding**

This work was partially supported by JSPS KAKENHI Grant No. JP18H02040, the MEXT Quantum Leap Flagship Program (MEXT Q-LEAP) Grant No. JPMXS0118067246, and JST CREST Grant No. JPMJCR19J4, Japan.